\documentclass{PoS}
\usepackage{bm}
\usepackage{amsmath}

\title{Tagged spectator DIS on a polarized spin-1 target}

\ShortTitle{Tagged spectator DIS on a polarized spin-1 target}

\author{\speaker{W. Cosyn},$^a$ M. Sargsian,$^b$ C. Weiss$^c$\\
       \llap{$ˆa$} Dept. of Physics and Astronomy, Ghent University,
Proeftuinstraat 86, B9000 Ghent, Belgium\\
\llap{$ˆb$} Department of Physics, Florida International University,
Miami, FL 33199, USA\\
\llap{$ˆc$} Theory Center, Jefferson Lab,
Newport News, VA 23606, USA\\
        E-mail: \email{wim.cosyn@ugent.be}, \email{sargsian@fiu.edu}, 
\email{weiss@jlab.org}}

\abstract{We discuss the process of deep-inelastic electron 
scattering (DIS) on the
polarized deuteron with detection of a nucleon in the nuclear fragmentation 
region (``spectator tagging'').
We cover (a) the general structure of the  semi-inclusive DIS cross section on a 
spin-1 target;    (b) the tagged  structure 
functions in the impulse approximation, where
deuteron structure is described by the $NN$ light-front wave function; (c) 
the extraction of free neutron structure through on-shell extrapolation
in the recoil proton momentum.
As an application 
we consider the extraction of
 the neutron spin structure function $g_{1n}$ through polarized electron 
scattering on the 
longitudinally polarized deuteron with
proton tagging and on-shell extrapolation. Such measurements would be possible 
at an
Electron-Ion Collider (EIC) with polarized deuteron beams and forward
proton detectors.}

\FullConference{XXIV International Workshop on Deep-Inelastic Scattering and Related Subjects\\
		11-15 April, 2016\\
		DESY Hamburg, Germany}

\begin{document}

\section{Introduction}

We consider the process of
tagged spectator DIS 
process on a deuteron target:
\begin{equation}
 e(l)+D(P_D) \rightarrow e'(l') + X (P_X) + N (p_N) \,,
 \label{eq:tagged_reaction}
\end{equation}
where a recoil nucleon is detected in the final state with low 
momentum, $\sim$	few 100 MeV in the deuteron rest frame.   Compared to 
more 
conventional 
inclusive deuteron DIS measurements, where only the scattered electron is 
measured, this 
reaction has the advantage that one has more control over the initial nuclear 
configuration of the deuteron target and the active nucleon (proton, neutron) 
is positively identified, thereby eliminating dilution from scattering on the 
other nucleon.  In inclusive measurements the struck 
nucleon has Fermi motion and one has to average over all nuclear 
configurations, as well as account for possible non-nucleonic components in the 
wave function.  
 Especially at high Bjorken $x$, the Fermi motion affects the uncertainties in 
the 
extraction of neutron structure functions and flavor separation of parton 
distribution functions.  Non-nucleonic components are suppressed in reaction 
(\ref{eq:tagged_reaction}) 
because of the deuteron target (isospin forbids $N\Delta$ components) and 
nucleon 
tagging ($\Delta\Delta$ contribution is eliminated).

The tagged spectator deuteron process has been measured in two pioneering 
experiments at Jefferson Lab (JLab).  The Deeps experiment 
\cite{Klimenko:2005zz} measured at proton momenta $300-600$ MeV, while the 
BONuS 
experiment \cite{Baillie:2011za,Tkachenko:2014byy} measured down to momenta of 
$70$ MeV.  Model calculations including the effect of final-state interactions 
(FSI) are available \cite{Cosyn:2010ux,Palli:2009it}.  A follow-up experiment is 
planned at the 12~GeV JLab upgrade~\cite{Bonus12} and another experiment will 
use the process to study the recoil momentum and isospin dependence of the EMC 
effect~\cite{Hen:2014vua} as the tagged nucleon offers control over the 
off-shellness of the struck nucleon and the density of the nucleus 
configuration.

Spectator tagging also enables the application of the so-called \emph{pole 
extrapolation}~\cite{Sargsian:2005rm}, where the cross section is extrapolated 
to the pole [$t=(p_D-p_N)^2=p_i^2=m_N^2$] of the propagator of the $t$-channel 
nucleon (see 
Fig.~\ref{fig:diagrams}a), and one can effectively probe on-shell neutron 
structure when tagging a proton and extracting the residue of the cross section 
at the pole.  Due to the small binding energy of the deuteron the actual 
extrapolation length into unphysical kinematics is very small and corresponds 
to almost zero but imaginary spectator momentum $p_N$ in the deuteron rest 
frame.  Pole 
extrapolation 
also has the advantage that the FSI contribution cancels as it does not have a 
pole for the on-shell exchange due to the loop momentum 
integration~\cite{Sargsian:2005rm}.  This makes pole extrapolation a very 
powerful, model-independent method of extracting neutron structure.  Recently, 
pole extrapolation was applied to the BONuS data and a surprising rise in the 
$F_{2n}/F_{2p}$ ratio at high $x$ was observed, albeit at sub-DIS values of 
$Q^2$~\cite{Cosyn:2015mha}.

Deuteron DIS with spectator tagging will be possible at a future EIC.  In such 
measurements, the spectator nucleon moves forward with $\sim$1/2 of the beam 
momentum and is detected with forward detectors.  Moreover, in a collider 
there is no target material. Compared to a fixed target 
setup, where the detectors have to be very close to the target and physically 
small (which limits their performance), this makes 
detection of the spectators more straightforward.  In particular, an EIC could 
enable tagged DIS with 
polarized deuteron beams.  These beams would be available in the Jefferson Lab 
EIC (JLEIC) figure-8 ring design~\cite{Cosyn:2014zfa,Cosyn:2016oiq} and allow 
for the measurement of neutron spin structure over a wide kinematic domain. The 
small 
extrapolation length for the tagged spectator process also means 
the $D$-wave admixture is very small and deuteron polarization is 
almost 100\% transferred to the nucleons. 

In 
these proceedings, we summarize the theoretical formalism for tagging on a 
polarized deuteron and show estimates for the vector polarized double spin 
asymmetry in kinematics accessible at an EIC.  The full formalism will be 
presented in Ref.~\cite{CMW} and the work presented here is part of a dedicated 
LDRD program at JLAB providing simulations for an EIC~\cite{LDRD}.


\vspace{-0.2cm}
\section{Formalism}
\vspace{-0.2cm}

To compute observables of reaction (\ref{eq:tagged_reaction}) with polarized 
initial beams, the 
general form of the SIDIS spin-1 cross section can be used~\cite{CMW}, as 
detecting a recoil 
nucleon 
can be considered SIDIS in the target fragmentation region.  The spin-1 target 
ensemble is described by a 3 by 3 density matrix 
$\rho_{\lambda\lambda'}$, which has 8 parameters (3 vector, 5 tensor 
polarization) \cite{Leader:2001gr}. 
The hadronic tensor is written as
 \begin{equation}
 W^{\mu\nu}= 
\sum_{\lambda'\lambda}\rho_{\lambda\lambda'}W^{\mu\nu}(\lambda',\lambda)\,,
\label{eq:hadronictensor_contracted}
\end{equation}
and obeys the transversality conditions $q_\mu W^{\mu\nu}=W^{\mu\nu} q_\nu=0$ 
and parity invariance constraints.  The cross section has 
41 structure functions 
each with an azimuthal modulation depending on the angle of the recoil nucleon 
and parameters of the density matrix~\cite{CMW}:
\begin{equation} 
\frac{d\sigma}{dxdQ^2d\phi_{l'}}=\frac{y^2\alpha^2_{\text{em}}}{Q^4(1-\epsilon)}
\left(\mathcal{F}_U+\mathcal{F}_S+\mathcal{F}_ 
T\right)d\Gamma_{P_N}\label{eq:fullcross}\,.
\end{equation}
The 5 upolarized (contained in $\mathcal{F}_U$) and 13 vector polarized 
(present in $\mathcal{F}_S$) terms have identical azimuthal and vector 
polarization dependence as the spin 1/2 case \cite{Bacchetta:2006tn}, while the 
23 structure functions contained in $\mathcal{F}_T$ are unique to the spin 1 
case and will be presented in detail elsewhere \cite{CMW}.  

\begin{figure}[htb]
\begin{center}
 \includegraphics[width=0.7\textwidth]{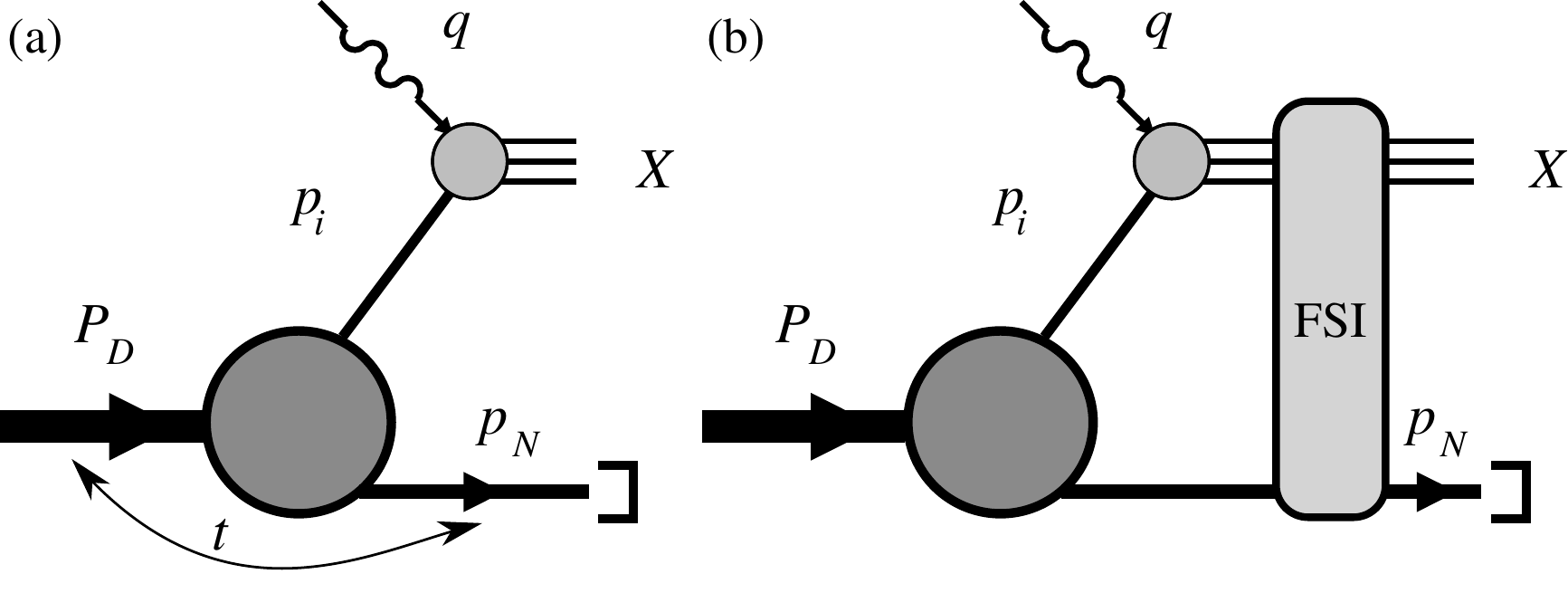}
 \end{center}
 \caption{
 \label{fig:diagrams}
Impulse approximation (a) and FSI (b) of the 
tagged spectator DIS process on 
a composite deuteron target.}
\end{figure}

A dynamical model for reaction (\ref{eq:tagged_reaction}), suitable 
for high-energy probes, is used where the nuclear structure is factorized from 
the 
probe--nucleon interaction~\cite{CMW}.  We show results in the impulse 
approximation (IA) depicted 
in 
Fig.~\ref{fig:diagrams}a and do not include the FSI
contribution of Fig.~\ref{fig:diagrams}b as it has different analytic 
properties 
compared to the IA and does not contribute at the on-shell nucleon pole.
In the IA, the hadronic tensor reads~\cite{CMW}
\begin{equation}\label{eq:htensorfinal}
 W^{\mu\nu}_{D}(\lambda',\lambda)=4(2\pi)^3 
\frac{\alpha_N}{2-\alpha_N} 
\sum_{i=U,z,x,y} W^{\mu\nu}_{N,i} 
\rho^i_D(\lambda',\lambda)\,,
\end{equation}
where $\alpha_N=2p_N^+/P_D^+$ is the recoil nucleon lightcone momentum 
fraction, the $W_{N,i}^{\mu\nu}$ depend on structure functions of the struck 
nucleon $F_{1N},F_{2N},g_{1N},g_{2N}$ evaluated at $\tilde{x}\approx 
x/(2-\alpha_N)$ and $Q^2$, and $\rho^i_D(\lambda',\lambda)$ are deuteron 
light-front densities depending on the deuteron light-front wave 
function.  In high-energy scattering it is natural to describe deuteron 
structure using
the light-front wave function, as in this formulation the off-shellness of
the $NN$ component remains finite in the high-energy limit, which is not the 
case 
in the instant form.  The light-front deuteron 
wave function 
$\Psi^D_\lambda(\bm k_f,\lambda_1,\lambda_2)$ has the same structure and 
rotational invariance properties as the non-relativistic wave function 
\cite{Frankfurt:1981mk,Keister:1991sb,Chung:1988my}, but differs in two 
important aspects: a) there are Melosh rotations $R_{fc}$ accounting for 
relativistic spin; b) it depends on a dynamical three-momentum $\bm k_f$ which 
is 
related to the kinematical momenta of the target deuteron and recoil nucleon 
$p_N$ in the following manner:
\begin{align}
 \frac{k_f^z}{E_{k_f}}=1-\alpha_N\,,
 \qquad\bm k_f^\perp = -\bm p_N^\perp + \frac{\alpha_i}{2}\bm P_D^\perp\,,
 \qquad E_k = \frac{m_N^2+\bm k_f^{\perp2}}{\alpha_N(2-\alpha_N)}\,.
 \label{eq:lf_k_mom}
\end{align}
The light-front wave function reads
\begin{equation}
 \Psi^D_\lambda(\bm 
k_f,\lambda_1,\lambda_2)=\sqrt{E_{k_f}}\sum_{\lambda'_1\lambda'_2}\mathcal{D}^{
\frac{1}{2}}_{\lambda_1\lambda'_1}[R_{fc}(k_{1_f}^\mu/m_N)]
\mathcal{D}^{\frac{1}{2}}_{\lambda_2\lambda'_2}[R_{fc}(k_{2_f}^\mu/m_N)]
\Phi^D_\lambda(\bm k_f,\lambda'_1,\lambda'_2)\,,\label{eq:D_lf_wf}
\end{equation}
where $\Phi^D_\lambda(\bm k_f,\lambda'_1,\lambda'_2)$ has the structure of 
a non-relativistic deuteron wave 
function, and $\mathcal{D}^{\frac{1}{2}}_{\lambda\lambda'}$ denote the Wigner 
D-matrices.

Using Eq.~(\ref{eq:htensorfinal}) it is possible to write all 41 structure 
functions that appear in Eq.~(\ref{eq:fullcross}) as a linear combination of 
the tensor and polarization components of 
$W^{\mu\nu}_{D}(\lambda',\lambda)$~\cite{CMW}.

\section{Results}

As an observable of interest in the tagged spectator process, we consider the 
double spin asymmetry for a longitudinally vector polarized target:
\begin{equation}
 A_\parallel = 
\frac{\sigma(++)-\sigma(+-)-\sigma(-+)+\sigma(--)}{
\sigma(++)+\sigma(+-)+\sigma(-+)+\sigma(--)} = 
\frac{F_{LS_L}}{F_{UU,T}+\epsilon F_{UU,L}}\,,
\end{equation}
where $\sigma(\lambda_e,\lambda_D)$ denotes the cross section for spin 
projections $\lambda_e=\pm \frac{1}{2}$,$\lambda_D=\pm 1$, averaged over 
the azimuthal angle of the recoil proton.  The $F_i$ are structure functions 
appearing in the cross section~\cite{CMW,Bacchetta:2006tn}, for which closed 
expressions have been obtained in the IA model~\cite{CMW}.  We can 
schematically write the structure function appearing in the nominator of 
$A_\parallel$ as a product of four components:
\begin{align}
 F_{LS_L}&=\{\text{kin. 
factors and deuteron polarization}\}\times \{ g_1(\tilde{x},Q^2), 
g_2(\tilde{x},Q^2) \} \nonumber\\
&\qquad\times \{\text{bilinear forms in deuteron radial wave function } U(k), 
W(k)\}\nonumber\\
&\qquad\times \{ \text{rel. spin factors resulting from Melosh transf. 
depending on } \bm{k}\}\,,
\label{eq:F_LSL_full}
\end{align}
where $U(k)\,[W(k)]$ is the deuteron radial $S$-wave [$D-$wave].  The 
$F_{UU,T}$ and $F_{UU,L}$ functions in the denominator have a similar form but 
are spin-independent so involve the nucleon structure functions 
$F_{1N},F_{2N}$ in the second component and omit the last 
component in Eq.~(\ref{eq:F_LSL_full}) (Details will be given in 
Ref.~\cite{CMW}).

\begin{figure}[htb]
 \begin{center}
  \includegraphics[width=0.6\textwidth]{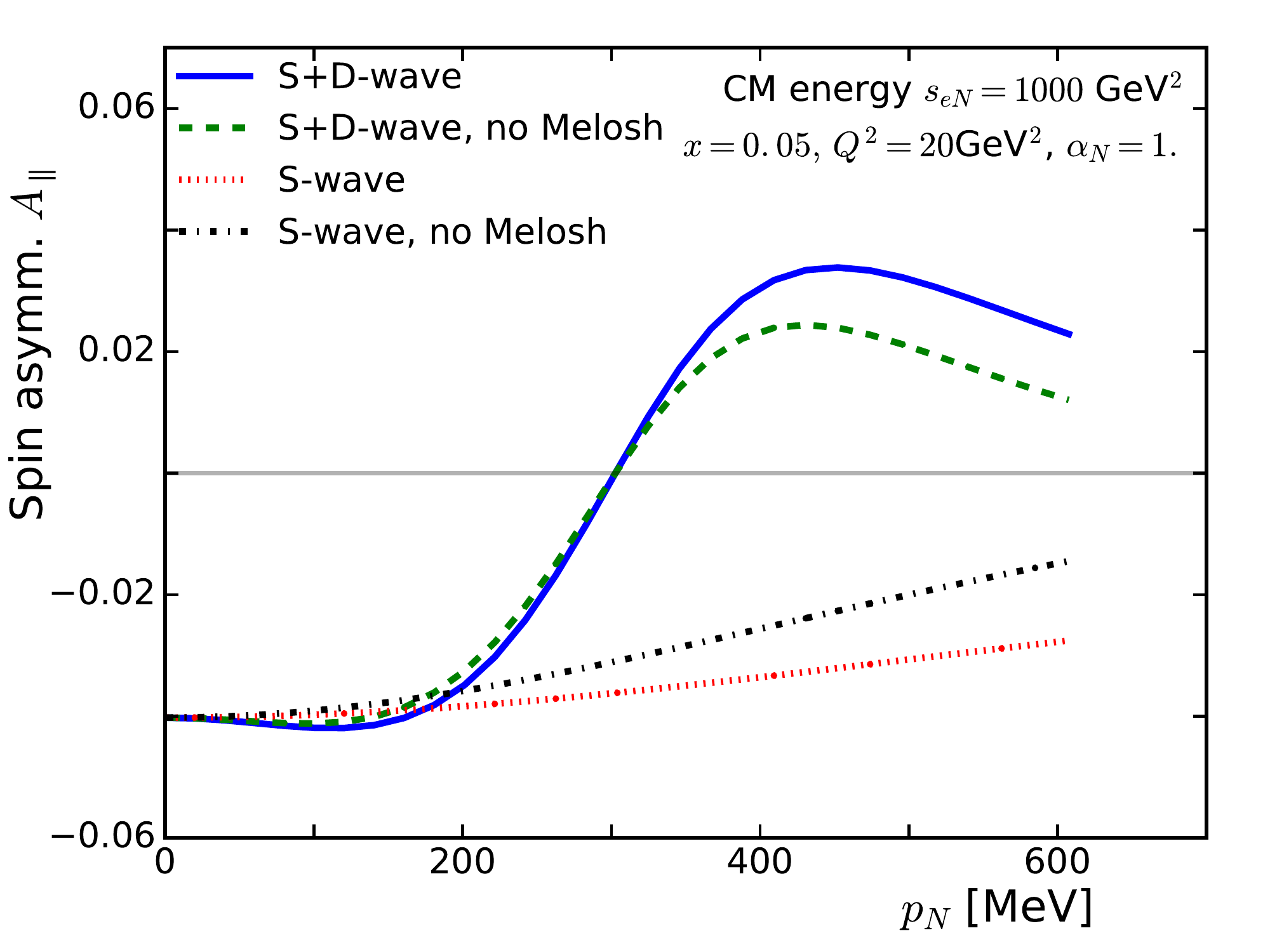}
  \caption{
  \label{fig:vecpol}
Spectator momentum dependence of the double spin asymmetry $A_\parallel$ as a 
function of the spectator proton momentum in the deuteron rest frame. Full blue 
curve 
includes both the $S$- and $D$-wave deuteron components and accounts for the 
Melosh rotations in the light-front wave function, dashed green curve 
only includes the deuteron $S$-wave component and accounts for the 
Melosh rotations in the light-front wave function.  The black dashed-dotted 
curve (only $S$-wave component) and red dotted curve ($S$- and $D$-wave 
components) do not include the Melosh rotations in the wave function. }
 \end{center}
 \vspace{-0.5cm}
\end{figure}

Pole extrapolation for $A_\parallel$ is straightforward as the nucleon pole  
$1/(m_N^2-t)^2$  in 
the cross sections, which appears in $A_\parallel$ through the radial deuteron 
wave function dependence [$U(k),W(k)$] of all structure functions 
[Eq.~(\ref{eq:F_LSL_full})],  cancels between the numerator and denominator.
Experimentally, the ratio has the advantage that systematic errors largely 
cancel.  When performing pole extrapolation on $A_\parallel$, one obtains the 
result
\vspace{-0.3cm}
\begin{equation}
 A_\parallel \quad \text{on-shell} \quad \sim D\,\frac{g_{1n}}{F_{1n}}\,,
\vspace{-0.2cm}
\end{equation}
where $D$ is the depolarization factor, which emerges from the kinematic 
factors in the cross section.  Then the on-shell $A_\parallel$ 
 together with data for $F_{1n}$ allows one to extract on-shell 
neutron spin structure.  Figure~\ref{fig:vecpol} shows a calculation of 
$A_\parallel$ as a function of recoil proton 
momentum  at kinematics accessible at an EIC and illustrates the 
influence of the deuteron D-wave and the Melosh rotations in 
Eq.~(\ref{eq:F_LSL_full}).  One clearly observes the influence of the 
$D$-wave (with largest size effects around the tensor force dominated region 
of 300-500 MeV) 
and also the influence of the Melosh rotations for higher recoil momenta.  Both 
effects 
effectively drop out towards the on-shell nucleon pole and the asymmetry is 
quite flat at recoil momenta $\leq 100$ MeV, making the extrapolation 
for the IA robust.
\vspace{-0.4cm}
\section{Summary}
\vspace{-0.3cm}
In sum, spectator tagging with polarized deuteron at EIC would enable precision
measurements of neutron spin structure over a wide kinematic range.
Further R\&D are needed to extend the theoretical framework (final-state 
interactions, nuclear shadowing effects) and realize the full potential of
this method.

\vspace{-0.3cm}
\acknowledgments
\vspace{-0.3cm}
This material is based upon work supported by the U.S. Department of Energy, 
Office of Science, Office of Nuclear Physics under contract DE-AC05-06OR23177.
\vspace{-0.3cm}

\bibliography{../../Articles/bibtexall}
\end{document}